\documentclass[10pt,journal,compsoc]{IEEEtran}
\ifCLASSOPTIONcompsoc
  \usepackage[nocompress]{cite}
\else
  \usepackage{cite}
\fi
\ifCLASSINFOpdf
\else
\fi

\usepackage{cite}
\usepackage{amsmath,amssymb,amsfonts}
\usepackage{algorithmic}
\usepackage{graphicx}
\usepackage{textcomp}
\usepackage{xcolor}
\def\BibTeX{{\rm B\kern-.05em{\sc i\kern-.025em b}\kern-.08em
    T\kern-.1667em\lower.7ex\hbox{E}\kern-.125emX}}
    
\usepackage{bm}
\usepackage{hyperref}

\newcommand{\sj}[1]{\textcolor{red}{#1}}

\newcommand{\model}{\texorpdfstring{T\MakeLowercase{o}PNN}{ToPNN}}
\newcommand{\app}{\texorpdfstring{C\MakeLowercase{o}ST\MakeLowercase{op}}{CoSTop}}

\begin{document}
%
\title{A Topic Guided Pointer-Generator Model for Generating Natural Language Code Summaries}
%
%
%
%

\author{Xin~Wang,
		Xin~Peng,~\IEEEmembership{Member,~IEEE,}
        Jun~Sun,
        Yifan~Zhao,
        Chi~Chen,
        and~Jinkai~Fan
\IEEEcompsocitemizethanks{\IEEEcompsocthanksitem X. Peng is the corresponding author.
\IEEEcompsocthanksitem X. Wang, X. Peng, Y. Zhao, C. Chen, and J. Fan are with the School of Computer Science and the Shanghai Key Laboratory of Data Science, Fudan University, Shanghai, China and Shanghai Institute of Intelligent Electronics \& Systems, China.
\IEEEcompsocthanksitem J. Sun is with the Singapore Management University, Singapore.}
}

%
%

\markboth{Journal of \LaTeX\ Class Files,~Vol.~14, No.~8, August~2015}%
{Shell \MakeLowercase{\textit{et al.}}: Bare Demo of IEEEtran.cls for Computer Society Journals}
%



\IEEEtitleabstractindextext{%
\begin{abstract}
Code summarization is the task of generating natural language description of source code, which is important for program understanding and maintenance.
Existing approaches treat the task as a machine translation problem (e.g., from Java to English) and applied Neural Machine Translation models to solve the problem.
These approaches only consider a given code unit (e.g., a method) without its broader context.
The lacking of context may hinder the NMT model from gathering sufficient information for code summarization.
Furthermore, existing approaches use a fixed vocabulary and do not fully consider the words in code, while many words in the code summary may come from the code.
In this work, we present a neural network model named \model~for code summarization, which uses the topics in a broader context (e.g., class) to guide the neural networks that combine the generation of new words and the copy of existing words in code.
Based on the model we present an approach for generating natural language code summaries at the method level (i.e., method comments).
We evaluate our approach using a dataset with 4,203,565 commented Java methods.
The results show significant improvement over state-of-the-art approaches and confirm the positive effect of class topics and the copy mechanism.
\sj{
}
\end{abstract}

\begin{IEEEkeywords}
code summarization, comment generation, program comprehension, deep learning, topic modeling
\end{IEEEkeywords}}

\maketitle

\IEEEdisplaynontitleabstractindextext

%
\IEEEpeerreviewmaketitle


%
%
%
%
%


\section{Introduction}
\IEEEPARstart{C}{ode} comments provide short summaries of source code in natural language.
They can help developers to achieve high degrees of program comprehension with both clarity and depth~\cite{kis08Comprehension}, and thus are important for program understanding and maintenance.
However, it is common that comments are missing, outdated, or mismatched~\cite{ESE05Maintenance, WCRE07Comment, ASE13AutoComment, SANER2015CloCom}.
Therefore, it is desirable that source code can be automatically summarized in a natural language based on code analysis.

Code summarization (a.k.a code comment generation) has long been researched in the software engineering community.
Template based approaches~\cite{ASE2010TowardsCommentsForJavaMethods, ICSE2011HighLevelActionMethod, SANER2017ObjectRelatedDescription, ICSME18Rule} rely on predefined rules or templates, and thus are limited in the types of summaries that can be generated.
IR (information retrieval) based approaches~\cite{WCRE2010TextSummarizationForSourceCodeIR,ICPC2014CODES,SANER2015CloCom} rely on similar code snippets to select or synthesize a comment, and thus are limited in the capability of learning from existing code.
With the availability of large source code repositories, data-driven approaches based on deep learning have largely overtaken template- and IR-based approaches~\cite{ACL2016CODENN, ICPC2018Deepcom, AAAI2018LiangCodeGRU, IJCAI2018SummCodeTransferredAPI}.
Inspired by the success of Neural Machine Translation (NMT)~\cite{NIPS2014Seq2Seq, CoRR17NMT} in natural language processing (NLP), some researches~\cite{ACL2016CODENN, ICPC2018Deepcom, ICSE2019LeClair} have applied NMT models for code summarization.
These approaches treat code summarization as a machine translation problem from a programming language (e.g., Java) to a natural language (e.g., English).
They use a sequence-to-sequence (Seq2Seq) model~\cite{NIPS2014Seq2Seq} with an encoder-decoder architecture and an attention mechanism to process sequence-based representations of source code (e.g., tokens) and generate natural language summaries word by word.
Recently, LeClair \emph{et al.}~\cite{ICSE2019LeClair} presented a code summarization approach that combines words from code with a flattened representation of code structure from AST (Abstract Syntax Tree) in a Seq2Seq model.

Similar to the process of human understanding, NMT requires comprehensive information for effective code summarization. Existing code summarization approaches only consider a given code unit (e.g., a method) without its broader context (e.g., a class).
This lacking of context may hinder the understanding of the code.
On the other hand, NMT based approaches generate a series of words from a fixed vocabulary based on estimated probabilities and do not fully consider the words in code.
While we agree with~\cite{ICSE2019LeClair} that the words in code are not important semantically (i.e., the behaviour of the code), we believe they are extremely important when it comes to code summarization and comprehension.
In fact, we observe that in the dataset used by~\cite{ICSE2019LeClair} 32.7\% of the words (after splitting by camel case and underscore) in the method comments can be found in the code.
Therefore, it is beneficial (if not necessary) to consider the words in code for code summarization.

In this work, we present a neural network model named \model~for code summarization.
\model~is designed based on our observation on the importance of code context and that of copying ``words'' from the source code for code summarization.
The basic idea is to use the topics in a broader context (e.g., class) to guide the neural networks that combine the generation of new words and the copy of existing words in code.
\model~includes a code topics encoder, a standard code tokens encoder, and a decoder.
The code topics encoder captures the semantic information of topics identified by a topic model. 
The decoder then combines the results of the code topics encoder and the code tokens encoder, and applies the pointer mechanism~\cite{ACL2017GetToThePoint}, which can point to and copy words from the source text, to choose between the decisions of generating new words and copying existing words from code.

Based on the \model~model we present an approach for generating natural language code summaries at the method level (i.e., method comments).
It includes an offline training phase and an online generation phase.
The offline training is based on a code corpus which includes a large number of commented methods.
We use topic mining techniques~\cite{NIPS2001LDA} to train a topic model and produce a topic distribution for each class. 
Then we extract the code snippets and comments of methods and use them together with their class topic distributions to construct a training corpus, as well as to train a \model~model.
To generate a summary for a code snippet (a method), we first use the topic model to infer a topic distribution for the class where the method resides and then provide the code snippet together with its related code topics for the \model~model to generate a summary.

To evaluate our approach we construct a dataset with 4,203,565 commented Java methods from 684,624 classes, which are collected from 8,254 open-source projects.
Each method in the dataset is associated with complete class information to facilitate class topic modeling.
To the best of our knowledge, this is \textit{the} biggest dataset for such study so far.
Based on the dataset, we evaluate our approach from three aspects.
First, we compare our approach with a standard Seq2Seq approach and a state-of-the-art approach Funcom~\cite{ICSE2019LeClair} using multiple metrics.
Second, we analyze the contribution of the key components of our model, i.e., the code topics encoder and the pointer mechanism of the decoder.
Third, we analyze the human feedback on the summaries generated by different approaches.

The comparison shows that \app~significantly outperforms a state-of-the-art approach Funcom~\cite{ICSE2019LeClair} and a standard Seq2Seq approach in terms of a range of NLP metrics including BLEU, ROUGE-L, METEOR, and CIDEr.
The contribution analysis shows that the code topics encoder alone provides significant improvement and the pointer mechanism provides significant improvement when combined with the code topics encoder.
The human study indicates that the summaries generated by \app~are better accepted than those generated by the two baseline approaches.



\section{Motivation}\label{sec:motivation}
\begin{figure}[t]
  \centering
  \includegraphics[width=0.75\linewidth]{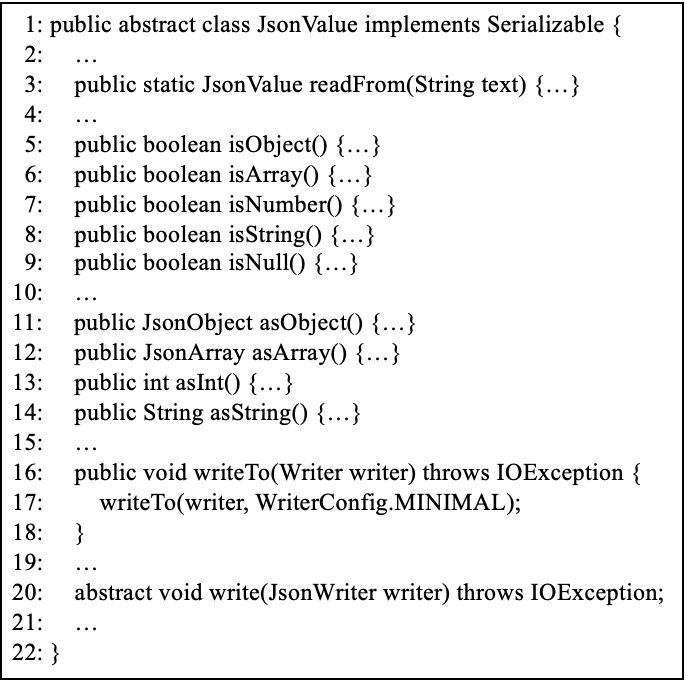}
  \caption{An Example for Code Topics} \label{fig:example-codeTopic}
\end{figure}

Our work is motivated by observation on the importance of broader code context and that of copying ``words'' from the source code for code summarization.

\subsection{Talk in Your Topic Areas}

Figure~\ref{fig:example-codeTopic} shows a method \textit{writeTo} in a class \textit{JsonValue}.
Its human written comment describes the method as ``writes the json representation of this value to the given writer in its minimal form without any additional whitespace''.
If we just read the code of the method, it is hard for us to understand its functionality, as the action in method name ``write'' is very general and the code does not include key concepts such as json format and value conversion.
If we use the code summarization approach proposed in~\cite{ICSE2019LeClair} and a standard Seq2Seq based approach to generate a summary based on the code of the method, we get ``writes the contents of the given code writer code to the given'' and ``writes the contents of this code byte buffer to the given writer'' as the summary respectively.
Writing the content of given writer to some given things and writing contents of an unknown buffer to the given writer are both far from being meaningful.

When a human developer tries to understand the method, usually he/she will first read the whole class and get a general context (e.g., json format and value conversion), which is important to the developer to understand what is to write through the writing operation of method \textit{writeTo}.

Base on the observation, our approach captures the class context of the method by using the topics of the class to guide the NMT model.
Using topic mining techniques we can get several relevant topics, 
for example json format, value conversion, and value validity checking. 
Guided by these topics our approach generates the summary ``writes the json representation of this object to the given writer''.

\begin{figure}[t]
  \centering
  \includegraphics[width=0.95\linewidth]{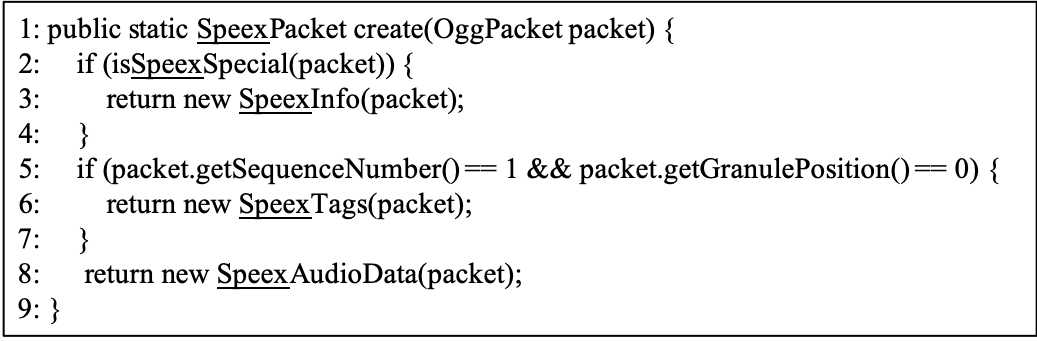}
  \caption{An Example for Code Copy} \label{fig:example-codeCopy}
\end{figure}

\subsection{Balance between Generating and Copying}
The method \textit{create} shown in Figure~\ref{fig:example-codeCopy} is described as ``creates the appropriate link speex packet'' in the human written comment.
The word ``speex'' refers to an audio compression format and rarely appears in code.
If we generate a summary from a fixed vocabulary, it is usually hard to generate the word ``speex''. 
If we use the approach in~\cite{ICSE2019LeClair} and a standard Seq2Seq based approach to generate a summary for it, we get ``creates a new tt codec packet tt instance from a specific jmf'' and ``create a codec for a packet'' respectively.
These two summaries are not meaningful and include invalid words.

A human developer can roughly understand the method as ``create speex packet'' based on its method name and return values.
Most of the words in this summary can be found in the code.

Our approach combines the strategies of generating new words and copying words from code.
For the method in Figure~\ref{fig:example-codeCopy} our approach generates the summary 
``creates a speex packet for the given ogg packet''.

\section{Background}\label{sec:background}

In this section, we present background techniques relevant to this work, including the Latent Dirichlet Allocation (LDA) model~\cite{NIPS2001LDA}, the Sequence to Sequence (Seq2Seq) model~\cite{NIPS2014Seq2Seq}, and the Pointer-Generator (Pointer-Gen) model~\cite{ACL2017GetToThePoint}.

\subsection{Topic Model}
Topic modeling is widely used in text mining and information retrieval to identify the topics in large amount of text corpus.
In this work, it is adopted to mine the latent topics in the context of methods (i.e., class topics) for guiding the code summarization.

The LDA model~\cite{NIPS2001LDA} is an unsupervised generative model in the form of a probabilistic graphical model for modeling collections of discrete data, which can be used for topic modeling. 
It has been proven effective in various areas, e.g., social media analysis~\cite{ictai2017LDASocialNetwork} and software engineering~\cite{kbse2007MineConcepts,icpc2014TopicModelCodeSummary}. The terms in the LDA model are defined as follows.
\begin{itemize}
	\item A \emph{word} is the basic unit of discrete data, which is represented as a vector $w$.
	\item A \emph{document} $d$ is a sequence of $N_{d}$ words, denoted as $d = (w_{d1},w_{d2},\ldots,w_{dN_d})$, where the $w_{dn}$ is the $n$-th word in $d$.
	\item A \emph{corpus} $D$ is a collection of $M$ documents.
\end{itemize}

An LDA model represents a corpus $D$ as random mixtures over $K$ latent \emph{topics}. Given a document $d$, the topic of a word $w_{dn}$, denoted as ${z_{dn}}$, is given by a multinomial distribution $z_{dn} \sim Multinomial(\theta_{d})$ where the $\theta_{d}$ is a random variable $\theta_{d} \sim Dirichlet(\alpha)$.
For each word $w_{dn}$ under a topic $z_{dn} = k$, the word is regarded as a multinomial variable given by the multinomial distribution for its topic $w_{dn} \sim Multinomial(\beta_{k})$.
Each $\beta_{k}$ in $\beta$ is a random variable $\beta_{k} \sim Dirichlet(\eta)$.

The only observed variables in the generative process are the words in $D$. The other variables are either latent variables (i.e., $\theta_{d}$, $z_{dn}$ and $\beta_{k}$) or hyper parameters (i.e., $\alpha$ and $\eta$).
The probability of a corpus is modeled as the following equation in~\cite{NIPS2001LDA}:
{\small \begin{equation}\label{equ:lda}
p(D|\alpha, \beta)=\prod _{d=1}^M \int p(\theta_{d}|\alpha) \left ( \prod _{n=1}^{N_d} \sum_{z_{dn}} p(z_{dn}|\theta_{d})p(w_{dn}|z_{dn},\beta) \right )d\theta_{d}
\end{equation}}

The parameters (i.e., $\theta_{d}$ and $\beta_{k}$) can be estimated by various method, such as Gibbs sampling~\cite{griffiths2002gibbs} and Variational Bayes (VB)~\cite{attias2000VB}.
In this work, we adopt the LDA to train a topic model for inferring the code topics of methods.
Each class is regarded as a document $d$, and each class token is regarded as a word $w_{dn}$.
Each class is thus seen as a document containing $n$ words, which then can be modeled by the topic model.
Afterwards, we could use the trained topic model to infer the class topics distribution, as well as the list of $N$ most related topics of the class.

\subsection{Sequence to Sequence Model}
Code summarization can be broadly regarded as a translation task which ``translates'' the code into its summarization.
The Seq2Seq model~\cite{NIPS2014Seq2Seq} is widely used for solving general sequence to sequence problems such as machine translation~\cite{CoRR2016Seq2SeqNMT} and summarization~\cite{EMNLP2015Seq2SeqSummarization}.

The Seq2Seq model uses an encoder to encode an input sequence (e.g., the sentence to translate) into a vector representation with fixed-length, and uses a decoder to generate an output sequence (e.g., the translation) based on this vector representation.
A Seq2Seq model is trained to maximize the probability of a correct output sequence given an input sequence.
Given an input sequence $x = x_1,x_2,\ldots,x_n$, the encoder reads one token from the input sequence at a time and computes the hidden state $h_t$ for each input time step $t$ as follows.

\begin{equation}\label{equ:seq2seq_h_x}
	h_t = f(x_t, h_{t-1})
\end{equation}
where $x_t$ is the current input token, $f$ is a nonlinear activation function, $h_{t-1}$ is the hidden state of the last input time step.

Based on the vector representation computed by the encoder, the decoder computes the state $s_i$ for each output time step $i$ as follows.
\begin{equation}\label{equ:seq2seq_s_y}
	s_i = f(y_{i-1}, s_{i-1})
\end{equation}
where $y_{i-1}$ is the output of the last output time step and $s_{i-1}$ is the state of the last output time step.

The decoder generates an output sequence $y = y_1,y_2, \ldots, y_m$, by generating one output token at a time until it outputs the token marking the end of the sequence. Each output token is generated as follows.
\begin{equation}\label{equ:seq2seq_p_y}
	p(y_i \mid y_1,y_2,\ldots,y_{i-1}) = g(y_{i-1}, s_i)	
\end{equation}
where the probability distribution $g$ is used to select the generated token $y_i$ based on the output $y_{i-1}$ of the last output time step and the current state $s_i$.

For a parallel corpus ${ \langle \mathcal{X}, \mathcal{Y} \rangle }_{s=1}^{S}$, the Seq2Seq model aims to optimize the following function:

\begin{equation}\label{equ:seq2seq_loss}
	L(\theta) = \sum_{s=1}^{S}logP(y_{(s)} \mid x_{(s)};\theta)
\end{equation}
where $\theta$ represents the parameters to optimize in the unit (e.g., Long Short-Term Memory (LSTM)~\cite{1997LSTM} unit, Gated Recurrent Unit (GRU)~\cite{corr2014ReferGRU}) used in the model, and $L(\theta)$ is a log-likelihood function. 
Standard optimization methods such as stochastic gradient descent (SGD)~\cite{IMS1952stochastic} and Adam~\cite{ICLR2015Adam} can help to maximize the likelihood and generate good target output sequence.

In the Seq2Seq model, an input sequence is encoded as a fixed-length vector, which makes it difficult to capture the most relevant information when dealing with long sequences.
The attention mechanism~\cite{ICLR2015Seq2SeqWithAttention} is proposed to solve this problem by dynamically selecting related input tokens when generating output tokens.

Before generating each output token $y_i$, the decoder computes a different context vector $c_i$ to represent the input tokens.
The $c_i$ consists of encoder outputs with different weights.
It also concatenates the context vector in computing current state $s_i$ and generating current output token $y_i$.
\begin{equation}\label{equ:att_s_y}
	s_i = f(y_{i-1}, s_{i-1}, c_i)
\end{equation}
\begin{equation}\label{equ:att_p}
	p(y_i \mid y_1,y_2,\ldots,y_{i-1}) = g(y_{i-1}, s_i, c_i)	
\end{equation}
\begin{equation}\label{equ:att_ci}
	c_i = \sum_{j=1}^{T}\alpha_{ij}h_j
\end{equation}
where the attention weight $\alpha_{ij}$ is used to computes $c_i$ with all the input tokens.
The $\alpha_{ij}$ is computed as:
\begin{equation}\label{equ:att_alpha}
	\alpha_{ij} = \frac{exp(e_{ij})}{\sum_{k=1}^{T}exp(e_{ik})}
\end{equation}
\begin{equation}\label{equ:att_e}
	e_{ij} = a(s_{i-1}, h_j)
\end{equation}
where $e_{ij}$ is a value to align input and output tokens to evaluate their relation.
There are many equations which can be used to compute value $\alpha_{ij}$.
The attention mechanism helps to measure the dependency between input tokens and output tokens, and thus improves the effectively of the Seq2Seq model in tasks like machine translation.

\subsection{Pointer-Generator Model}
As we show in Section~\ref{sec:motivation}, code summarization often requires us to copy words from the input.
To achieve that, we adopt the Pointer-Gen model~\cite{ACL2017GetToThePoint} in this work, which is a hybrid network that augments the traditional Seq2Seq model with the attentional mechanism.
It offers the ability to copy words from the input sequence and retains the competence to produce novel words at the same time, which is particularly important for code summarization.

A Pointer-Gen model facilitates copying words from the input sequence via a \emph{pointing mechanism} whilst generating words from a predefined vocabulary via a \emph{generator}. In particular, a Pointer-Gen model computes the attention weight $a_{ij}$ on the input sequence and the context vector $c_{i}$. For each output step $i$, the Pointer-Gen model uses the sigmoid function $\sigma$ to calculate a \emph{generation probability} $p_{gen} \in [0,1]$ as a soft switch for balancing between generating and copying:
\begin{equation}\label{equ:point_gen_p_gen}
p_{gen} = \sigma(w_{c_i}^Ic_i+w_{s}^Is_i+w_{y}^Iy_{i-1}+b_{ptr})
\end{equation}
where $s_i$ is the decoder state, $y_{i-1}$ is the decoder input (i.e. the output of last step), and $w_{c_i}$, $w_{s}$, $w_{y}$ and scalar $b_{ptr}$ are trainable parameters.

In order to decode, the decoder calculates the probability of the final distribution based on $p_{gen}$:
\begin{equation}\label{equ:point_gen_p_w}
P(w) = p_{gen} P_{vocab}(w) + (1-p_{gen})\sum_{i:w_{i}=w}a_{ij}
\end{equation}
where $P_{vocab}$ is the probability to generate a word from the fixed vocabulary which is produced based on the hidden state of the decoder and the context vector, and $a_{ij}$ is the attention weight on the input sequence.
Since the word $w$ (weighted by $a_{ij}$ which is introduced by the input sequence) may be out-of-vocabulary (OOV), the final distribution produced by the Pointer-Gen model is over the \emph{extended vocabulary} which combines the words in the predefined vocabulary and the vocabulary introduced by the input sequence.
Note that in our context, we use the Pointer-Gen model to solve the OOV problem in code summarization.

\section{\model~Model}\label{sec:model}
\begin{figure}[t]
  \centering
  \includegraphics[width=0.9\linewidth]{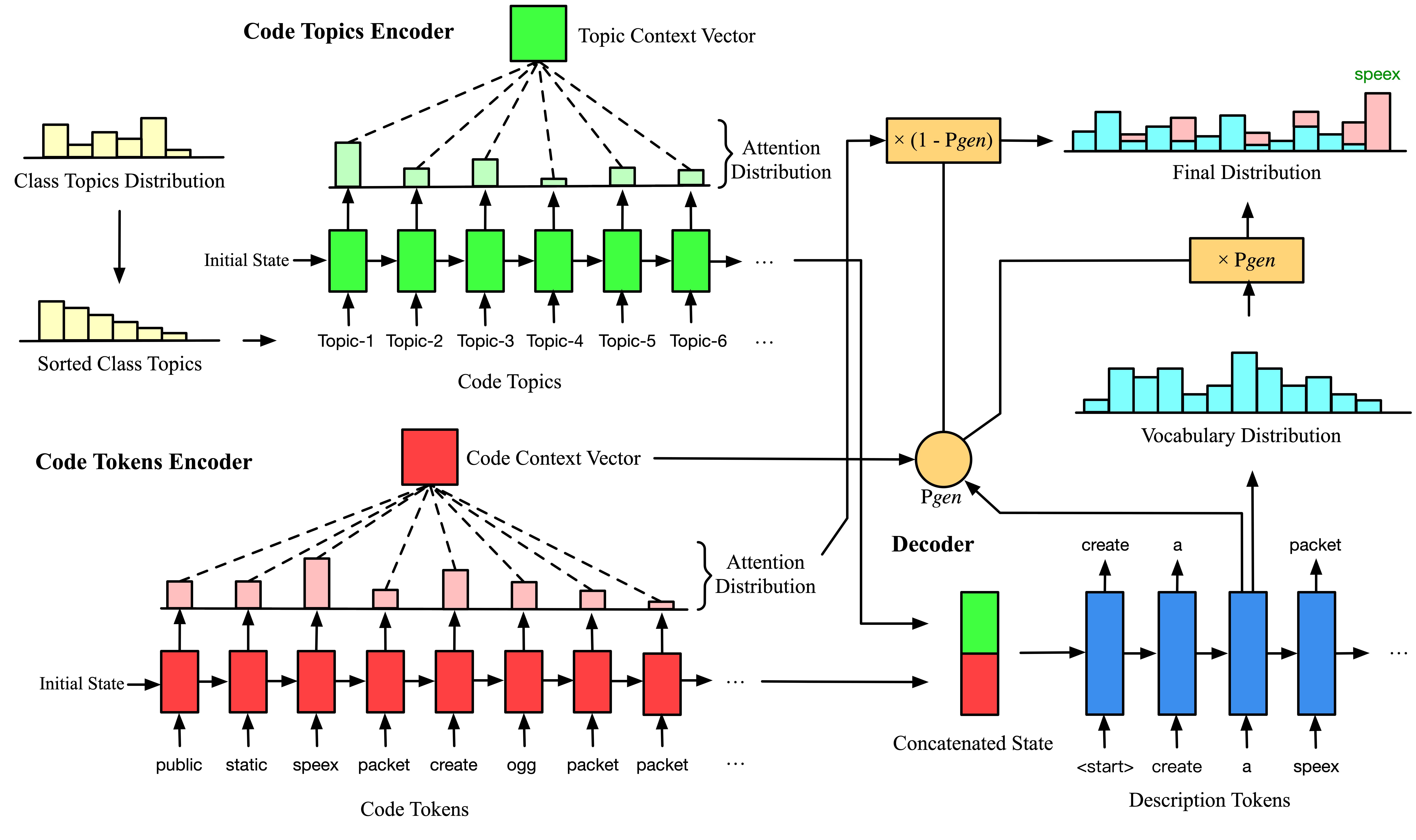}
  \caption{Architecture of \model}
  \label{fig:ourmodel}
\end{figure}

In this section, we present details of the novel neural network model named \model~which we design specifically for code summarization. \model~is designed based on our observation on the importance of topics and that of copying ``words'' from the source code. The high-level design of \model~is shown in Figure~\ref{fig:ourmodel}.
%
%
%
%
\model~consists of three components, i.e., a code topics encoder, a code tokens encoder and a decoder which incorporates a pointer mechanism. The training set of \model~contains triples of the form $(code$-$topics$, $code$-$tokens$, $description$-$tokens)$ where $code$-$topics$ are the output of the topic model, i.e., a list of $N$ most related topics; $code$-$tokens$ are the results of tokenizing a method; and $description$-$tokens$ are the results of tokenizing the corresponding code summary (i.e., method comment).
Once the model is trained, in the generating phase, \model~takes as input a pair of $code$-$topics$ and $code$-$tokens$ and output a code summary. \\

\noindent \emph{The code topics encoder} is designed to capture the semantic information in the topics identified by a topic model. It guides the generation of code summary by embedding the identified topics into a vector. The code topics are encoded as a sequence as follows.
\begin{equation}
	X^{top}=(x_1^{top},x_2^{top},\ldots,x_n^{top})
\end{equation}
where $n$ is the length of the sequence and $x_i^{top}$ where $1 \leq i \leq n$ represents a code topic. Note that instead of treating code topics as a bag-of-words, they are encoded as a sequence where the position of each code topic reflects the rank of the topic.
The code topics encoder embeds the topic sequence into a topic context vector, denoted as $V^{top}$, and computes its states $h^{top}_t$ by GRUs as follows.
\begin{equation}
	V^{top} = Embed(x_1^{top},x_2^{top},\ldots,x_n^{top})
\end{equation}
\begin{equation}
	h_t^{top}=GRU^{top}(v_t^{top},h_{t-1}^{top})
\end{equation}
where $Embed$ is the embedding layer to embed code topics whose result $V^{top}$ is a sequence $(v_1^{top}, v_2^{top}, \cdots, v_n^{top})$;  $GRU^{top}$ is the GRU unit; $v_t^{top}$ is a vector representation of $x_t^{top}$, and $h_{t-1}^{top}$ is previous state.
Note that GRU processes each topic in the sequence one-by-one, i.e., the new state $h_t^{top}$ is computed based on the top $v_t^{top}$ and the previous state $h_{t-1}^{top}$ which encodes information of all previous topics in the sequence.
The reason of using a topic sequence instead of a bag-of-words is that we think it is justified to consider the most dominant topic first and then the others. 
After processing the last topic, the code topics encoder collects and encodes the information of all the topics in the sequence.
The last state of the code topics encoder is denoted as $h^{top}_{n}$. \\

\noindent \emph{The code tokens encoder} is designed to capture the semantic information of a given method which is represented as code tokens.
The code tokens take the form of a sequence as follows.
\begin{equation}
	X^{cod}=(x_1^{cod},x_2^{cod},\ldots,x_m^{cod})
\end{equation}
where $m$ is the length of the sequence and $x_i^{cod}$ where $1 \leq i \leq m$ represents a code token.
Similarly, the code tokens encoder embeds the sequence of code tokens into a code context vector denoted as $V^{cod}$ and computes its states $h_t^{cod}$ using GRUs as follows.
\begin{equation}
	V^{cod} = Embed(x_1^{cod},x_2^{cod},\ldots,x_m^{cod})
\end{equation}
\begin{equation}
	h_t^{cod}=GRU^{cod}(v_t^{cod},h_{t-1}^{cod})
\end{equation}
where $Embed$ is the embedding layer to embed code tokens, $GRU^{cod}$ is the GRU unit, $v_t^{cod}$ is the vector representation of $x_t^{cod}$, and $h_{t-1}^{cod}$ is the state of last code token at time step $t-1$.
After processing the last code token, the code tokens encoder collects and encodes the information of all the code tokens in the sequence. \\

\noindent \emph{The decoder} is designed to generate code summary token-by-token based on the output from the encoders.
The decoder also uses GRUs.
The initial state of the decoder is set to be $[h^{top}_{n};h^{cod}_{m}]$ rather than $h^{cod}_{m}$, as we would like to use the topics to guide the generation of code summary. As a result, the initial state of the decoder combines ``semantic'' information of both the code topics and the code tokens.
In order to capture the importance of each code topic and code token in the code-tokens encoder, 
we use an attention mechanism to compute the attention weights of each code topic and code token of the encoders.
The new weighted context vectors with attention are as follows.
\begin{equation}
	c_t^{top}=\sum_{j=1}^n\alpha_{tj}^{top}f^{top}(h_j^{top})
\end{equation}
\begin{equation}
	c_t^{cod}=\sum_{j=1}^m\alpha_{tj}^{cod}f^{cod}(h_j^{cod})
\end{equation}
where $\alpha_{tj}^{top}$ (and $\alpha_{tj}^{cod}$) is the attention weight of $j$-th code topic (and code token) for the current step; $h_j^{top}$ and $h_j^{cod}$ are the states of the code topics encoder and code tokens encoder at $j-$th step respectively; $f^{top}$ and $f^{cod}$ denote dense layers for keeping the same dimension for decoder; $c_t^{top}$ and $c_t^{cod}$ are the context vector of code topics and code tokens respectively. A description token is generated from a predefined vocabulary at a time step $t$ as follows.
\begin{equation}
	h_t^{des} = GRU([c_t^{top};v_{t-1}^{des};c_t^{cod}],h_{t-1}^{des})
\end{equation}
\begin{equation}
	P_{vocab}(w) = softmax(W_{vocab}^Th_t^{des}+b_{vocab})
\end{equation}
where $h_{t-1}^{des}$ is the state of the previous description token at time step $t-1$; $v_{t-1}^{des}$ is the output vector of description token at time step $t-1$; $W_{vocab}^T$ and $b_{vocab}$ are trainable parameters and $P_{vocab}$ is the distribution of description token $w$ over the predefined vocabulary.

Since it is better to copy from the code tokens (such as method names or variable names) in many situations, a pointer mechanism is adopted to achieve that. A soft switch named $p_{gen}$ is computed to represent the probability of generating a description token from the predefined vocabulary, and $(1-p_{gen})$ is computed to represent the probability of generating a description token from the code tokens.
As a result, the generation process is as follows.
\begin{equation}
	P(w)=p_{gen}P_{vocab}(w)+(1-p_{gen})\sum~_{i:w_i=w}^l\alpha~_i^{cod}
\end{equation}
where $P(w)$ is the extended distribution; $w_i$ is the $i$-th description tokens to generate; and $l$ is the length of the code summary.

The likelihood function of \model~to optimize is:
\begin{equation}
	L(\theta) = \sum_{s=1}^{S}logP(Y^{des}_{(s)} \mid X^{top}_{(s)}; X^{cod}_{(s)} ;\theta)
\end{equation}
where $X^{top}_{(s)}$ and $X^{cod}_{(s)}$ are the code topics and the code tokens of $s$-th training instance respectively.
The $Y^{des}_{(s)}$ is the correct code summary of $s$-th training instance with length $l$:
\begin{equation}
	Y^{des}=(w_1^{des},w_2^{des}\ldots,w_l^{des})
\end{equation}
and $\theta$ represents trainable parameters.

\section{\app~Approach}\label{sec:approach}
\begin{figure}[t]
  \centering
  \includegraphics[width=1\columnwidth]{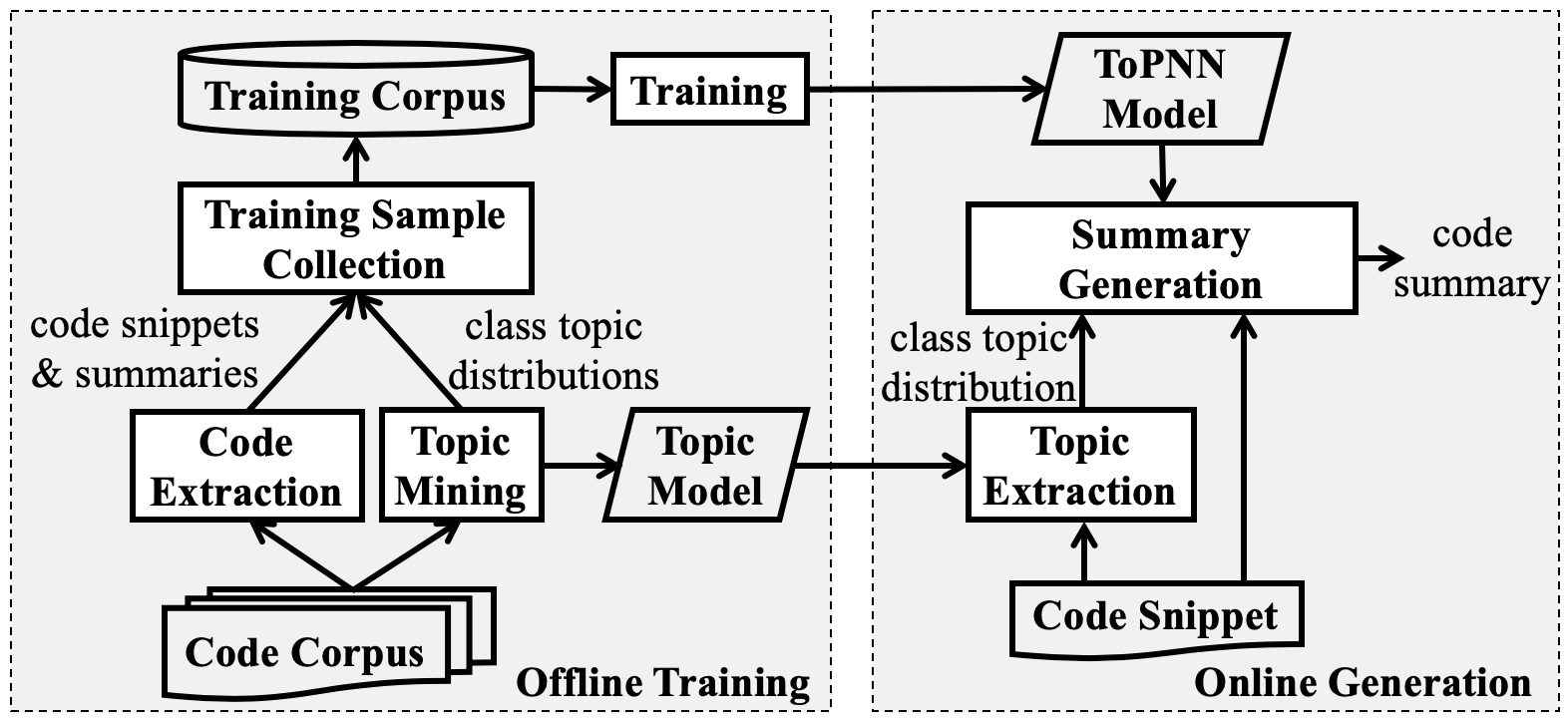}
  \caption{Overview of the \app~Approach} \label{fig:overview}
\end{figure}

In this section, we present our approach for code summarization, named \app, which is based on the \model~neural network model discussed in Section~\ref{sec:model}. An overview of \app~is shown in Figure~\ref{fig:overview}, which includes an offline training phase and an online generation phase. The former produces a topic model and a \model~model, which are used to generate code summaries in the generation phase.

The input of the offline training phase is a code corpus which includes a large number of commented methods.
To collect training instances, we extract code snippets (methods) and the corresponding code summaries (method comments) from the code corpus.
Meanwhile, we use topic mining techniques~\cite{NIPS2001LDA} to train a topic model and produce a topic distribution for each class. 
Note that we treat each class without the comments as a document during topic mining. 
The extracted code snippets and their summaries together with their class topic distributions are used to construct a training corpus.
The training corpus is then used to train a \model~model.

The input of the online generation phase is a code snippet (a method). 
To generate a code summary for the code snippet, we first use the topic model to infer a topic distribution for the class where the method resides. 
The code snippet together with its class topic distribution is then provided for the \model~model to generate a code summary.

\subsection{Training Corpus Construction}
As described in Section~\ref{sec:model}, a training instance of \model~is a triple, i.e., $(code$-$topics$, $code$-$tokens$, $description$-$tokens)$.
To construct the training instances, we collect a large corpus of open-source projects from GitHub and obtain 8,254 projects which are created during the period of 2010 to 2016. Note that only projects with more than 50 stars are selected for quality control. We adopt the criterions in~\cite{ICSE2019LeClair} to systematically extract methods and comments in the above projects as follows.
First, we extract and identify the methods which have non-trivial comments.
Second, for each identified method, we extract the corresponding JavaDoc comment~\cite{sigdoc1999KramerJavaDoc} (identified by ``/**''), which is the comment marked as the description of the code.
Following a common practice (e.g., \cite{ICSE2019LeClair}), which assumes that the first sentence of JavaDoc comment summarizes the method functionalities, we select the first sentence (identified based on the first ``.'' or ``\textbackslash n'') as the summary for the method.
Afterwards, only the comments written in English are retained using the language detection library ported from Google's language-detection.
Finally, we remove comments which include ``generated by'' to filter auto-generated comments.

In total, we obtain 4,203,565 pairs of method and comments. 
For each method, we further identify the containing class. 
We identify 684,624 classes in total. 
We then construct the training instances based on the pairs and the corresponding classes. 
For each class, since the topic model is trained to infer the top distribution of the source code, we remove all the comments in the class. 
For each method, we remove the comments in its body to reduce the noise. 
Afterwards, we tokenize each class, method and comment based on whitespace, split the tokens on camel case and underscore (if any), remove non-alpha tokens (e.g., ``='', ``)'', ``('', ``\}'', ``\{'') in the tokens, and convert the tokens into lowercase ones.
After the tokenization of the method and comment, we have $code$-$tokens$ and $description$-$tokens$ respectively.
Note that this is a standard process which is adopted by previous works~\cite{ICSE2019LeClair}.
In addition, for training the topic model using LDA~\cite{gensim}, we remove the general words in the class tokens, i.e., the Java keywords.
We further remove those tokens which are too short or too long, i.e., those class tokens with a length less than 2 or more than 15 (the parameters are the default setting of standard Gensim~\cite{gensim} library interface).
The topic model is trained on the corpus of all the processed classes in the training set (each class as a document, in the form of class tokens), and then used for inferring $code$-$topics$.
We obtain the most relevant topics of each method based on the containing class, i.e., by inputting the class tokens to the trained topic model.
Finally, we obtain the $(code$-$topics$, $code$-$tokens$, $description$-$tokens)$ triples as the training instances of \model.

\subsection{Model Training}
\begin{figure}[t]
  \centering
  \includegraphics[width=0.7\linewidth]{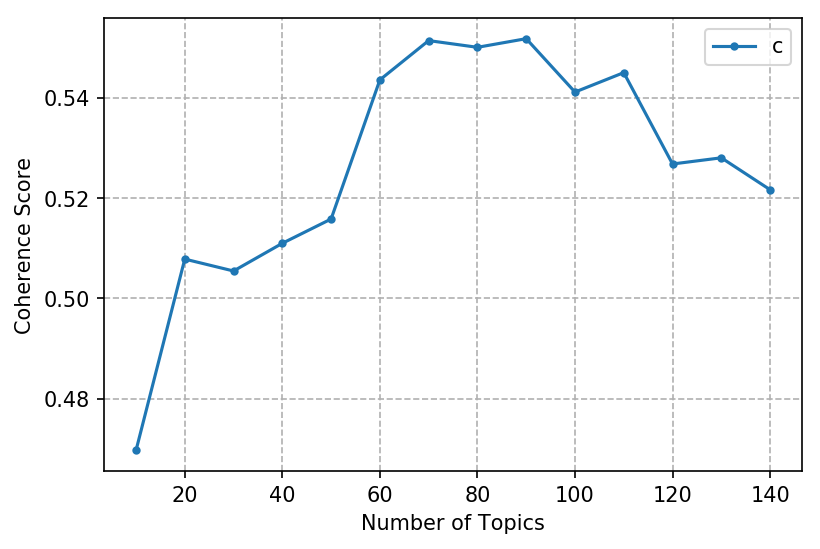}
  \caption{Coherence Score ($C_v$) for LDA Given Different Number of Topics}
  \label{fig:lda_k_cv}
\end{figure}

\noindent \emph{Topic Model Training}
To train the topic model, we need to find an ``optimal'' number of topics. 
To do that, we computes the coherence score~\cite{HLT-NAACL2010NewmanTopicCoherence}, denoted as $C_v$, for the LDA models trained with different number of topics. This is inspired by R{\"{o}}der \emph{et al.}~\cite{WSDM2015TopicCoherenceMeasures}.
The number, denoted as $K$, is set to be 10 initially, and incremented by 10 each time. For each value of $K$, we train the topic model and compute the $C_v$ coherence score. The result is shown in Figure~\ref{fig:lda_k_cv}. 
It can be observed that two peaks of almost the same coherence score are obtained and the first one comes when $K$ is $70$.
We thus set $70$ to be our number of topics.
Our topic model is implemented based on Gensim~\cite{gensim} (i.e., an open-source library which supports modern statistical algorithms).
Gensim supports online LDA parameter estimation based on work by Hoffman \emph{et al.} ~\cite{NIPS2010OnLineLearningLDA}.

\noindent \emph{\model~Training}
Once the training instances are collected, we train the \model~model.
The parameters of \model~are list as follows.
The hidden size of GRU cell is 256 in encoder and 512 in decoder as it uses the concatenated state as initial state;
the embedding dimensions for code topics, code tokens, and description tokens are 10, 100, 100, respectively; and the batch size is 128.
Following~\cite{ICSE2019LeClair}, we set the hyper parameters based on evidence gathered from trial experiments.
We train \model~for 10 epochs, and after each epoch, we estimate the BLEU~\cite{ACL2002BLEU} metric on the validation set.
Finally, we choose the trained \model~in the epoch with the best BLEU metric as our model.
\model~is implemented based on the Keras 2.2.4 (i.e., an open-source deep learning library) and uses the Adam optimizer with learning rate to be 0.001 to tune the  parameters.
\model~is trained on the server with 2 Intel Xeon E5-2666v3 CPUs, 4 GTX 1080Ti GPUs, and 64G RAM.

\subsection{Code Summarization}
Given a code snippet (i.e., a method) as input, \app~aims to generate the code summary of this code snippet.

First, the class which contains the method is identified. Next, all the comments in the class are eliminated and the remaining source code are tokenized into class tokens, which are then fed into the topic model to infer the topic distribution. The result is a list of top-N code topics that are most relevant to the class. Then all the comments in the body of the method are eliminated and the remaining source code are tokenized into code tokens. Afterwards, the code topics and code tokens are fed into the code topics encoder and code tokens encoder respectively. Finally, the decoder of \model~generates a code summary based on the topic context vector, the code context vector and previously generated description tokens.


\section{Evaluation}
To evaluate the effectiveness of \app, we conduct experimental studies to answer the following research questions.
\begin{itemize}
	\item \textbf{RQ1 (Performance Comparison)} How is the performance of \app~compared with state-of-the-art code summarization approaches in terms of different metrics?
	
	\item \textbf{RQ2 (Contribution of Components)} How does each component contribute to the performance of \app?

    \item \textbf{RQ3 (Human Feedback)} How are the summaries generated by different approaches viewed by developers?	
		
\end{itemize}

\subsection{Baselines, Code Corpus, and Metrics}\label{sec:bcm}
We compare our approach with a state-of-the-art approach Funcom~\cite{ICSE2019LeClair} and a standard Seq2Seq approach.
\textbf{Funcom} approach is developed by LeClair \emph{et al.}~\cite{ICSE2019LeClair} using an attentional Seq2Seq model to generate code summaries. It uses a structure-based traversal (SBT) method~\cite{ICPC2018Deepcom} to traverse AST of source code to convert it into a sequence of flatten structural tokens. Then, it uses two encoders (one for textual tokens and one for flatten structural tokens) to combine textual and structural information for its decoder to generate code summaries.
\textbf{Seq2Seq} approach is implemented based on Seq2Seq model~\cite{NIPS2014Seq2Seq} which consists of an encoder, a decoder, and attention mechanism. It takes a sequence of source code tokens as input and outputs code summaries. 
For both approaches we use the implementations provided by~\cite{ICSE2019LeClair}.
For Funcom we implement a processor to convert source code to SBT-AO (SBT with AST only) based on the description in~\cite{ICSE2019LeClair} and concatenate source code tokens together.

All the experimental studies are conducted based on our code corpus described in Section~\ref{sec:approach}.
The code corpus is split \emph{by project} into training set, validation set and test set.
``By project" here means that the projects are randomly divided into three disjoint sets in proportion: 90\% of the them are training set, 5\% are validation set, and 5\% are test set.

All the models are implemented and trained based on same server as \app.
For each model, the hidden size of GRU cell is set to 256 which is following ~\cite{ICSE2019LeClair};
and the batch size is set to 128.

We use the following four metrics, which are widely used in code summarization and related NLP tasks (e.g., machine translation, document summarization), to measure the quality of the generated summaries.

\textbf{BLEU}. BLEU~\cite{ACL2002BLEU} is widely used for automatic evaluation of machine translation. It is calculated by comparing the $n$-grams of the candidate translation with the $n$-grams of the reference translation and counting the number of matches in a position independent way. The more the matches, the better the candidate translation is.

\textbf{ROUGE-L}. Similar to BLEU, ROUGE~\cite{bookTextSumm2004ROUGE} also compares the $n$-grams of the candidate sentence with the $n$-grams of the reference sentence. It additionally computes $n$-gram based recall for the candidate sentence with respect to the references. ROUGE-L is the ROUGE version that computes an F-measure with a recall bias using longest common subsequence.

\textbf{METEOR}. METEOR~\cite{ACL2005METEOR} also computes the F-measure based on matches. At the same time, it resolves word-level correspondences in a more sophisticated manner, using exact matches, stemming and semantic similarity.

\textbf{CIDEr}. CIDEr~\cite{CVPR2015CIDEr} is a consensus based metric that considers the importance of words. It assigns weights for $n$-grams based on TF-IDF analysis and calculates the cosine similarity between the candidate sentence and the reference sentence based on the $n$-grams.

In our studies, we use NLTK (an NLP toolkit)~\cite{09NLTK} to calculate BLEU and the implementation by Chen \emph{et al.}~\cite{COCO} to calculate the other three metrics.
Following the common practice, for BLEU and CIDEr we calculate the values with 1-4 grams (i.e., $n=1, 2, 3, 4$) and take the average as the value of the metric.

\subsection{Performance Comparison (RQ1)}\label{sec:rq1}

We empirically observe that the code topics beyond the top 10 are less related to the code snippet, so we choose the length of code topics $N$ as 10 in this study.
The impact of the length of code topics on the performance of \app~will be discussed in RQ$_2$.

Table~\ref{table:performance} shows the results of the performance of \app~and baseline approaches on the test set which contains 149,520 test instances.
It can be seen that \app~achieves the best performance across all the metrics.
As the results show, in terms of BLEU metric, Seq2Seq achieves about 18.5 BLEU, Funcom achieves about 18.2 BLEU, and \app~achieves 19.5 BLEU which outperforms the other two approaches by 1.0 and 1.3, respectively.
In contrast, Funcom achieves an improvement of 0.2 BLEU over Seq2Seq as reported in~\cite{ICSE2019LeClair}.
Taking the other three metrics into consideration, \app~outperforms Seq2Seq (by 2.7 on ROUGE-L, 1.0 on METEOR, and 0.20 on CIDEr) and Funcom (by 1.9 on ROUGE-L, 1.0 on METEOR, and 0.17 on CIDEr) also.
The improvement of \app~is greater compared with the improvement of Funcom over Seq2Seq (by 0.8 on ROUGE-L, 0 on METEOR, and 0.03 on CIDEr).
In summary, the improvement of \app~is about 5.4-7.1\%, 4.8-6.9\%, 5.3-5.3\%, and 8.8-10.5\% over the two baseline approaches in terms of BLEU, ROUGE-L, METEOR, and CIDEr, respectively.

One extra observation is that the BLEU metric of Funcom is lower than Seq2Seq (by 0.3).
One possible reason is that the code corpus is different from that in~\cite{ICSE2019LeClair} and thus the data distribution is different.

\begin{table}[t]
\centering
\caption{Evaluation of Performance over \app~and Baseline Approaches (best scores are in boldface)}
\label{table:performance}
\setlength{\tabcolsep}{1.5mm}{
\begin{tabular}{|c|c|c|c|c|c|}
\hline
\textbf{Method} & \textbf{BLEU} & \textbf{ROUGE-L} & \textbf{METEOR} & \textbf{CIDEr} \\ 
\hline
Seq2Seq & 18.5 & 38.9 & 18.9 & 1.90 \\
Funcom & 18.2 & 39.7 & 18.9 & 1.93 \\
\app~($N$=10) & \textbf{19.5} & \textbf{41.6} & \textbf{19.9} & \textbf{2.10} \\
\hline
\end{tabular}}
\end{table}

\subsection{Contribution of Components (RQ2)}

To evaluate the contributions of each component of \app, we implement two variants of it based on its components:
\textbf{\app-Pointer-Only (\app-PO)},
which is a variant of \app~without the code topics encoder;
\textbf{\app-Topic-Only (\app-TO)},
which is a variant of \app~without the pointer mechanism in the decoder.
For each variant we follow the same process to train and test the model (see Section~\ref{sec:bcm}).
Note that the Seq2Seq approach in Section~\ref{sec:rq1} corresponds to a variant of \app~without the code topics encoder and the pointer mechanism in the decoder.

Table~\ref{table:component_performance} shows the results of the performance of Seq2Seq, \app-PO, \app-TO~($N$=10), and \app~($N$=10). 
Compared with the basic Seq2Seq approach, \app-PO performs better on ROUGE-L, METEOR, CIDEr, but worse on BLEU. 
The reason is that the pointer mechanism here is used to copy words from input in programming language which may not exactly match the words in evaluation dictionary.
\app-TO ($N$=10) performs better on all the four metrics.
Combining both the code topics encoder and the pointer mechanism, \app~($N$=10) outperforms all the variants on all the four metrics, especially ROUGE-L.
When compared with \app-TO~($N$=10), \app~($N$=10) performs better on all the four metrics, by 0.7 on BLEU, 1.6 on ROUGE-L, 0.7 on METEOR, and 0.17 on CIDEr.
When compared with \app-PO, \app~($N$=10) performs better on all the four metrics, by 2.2 on BLEU, 1.2 on ROUGE-L, 0.8 on METEOR, and 0.15 on CIDEr.
The results show that both components contribute to the approach, and the code topics encoder really boosts the overall approach, including both the basic Seq2Seq model and the pointer mechanism, indicating the importance of topic guidance.

\begin{table}[t]
\centering
\caption{Evaluation of Performance over \app~and Component Methods (best scores are in boldface)}\label{table:component_performance}
\setlength{\tabcolsep}{1.5mm}{
\begin{tabular}{|c|c|c|c|c|c|}
\hline
\textbf{Method} & \textbf{BLEU} & \textbf{ROUGE-L} & \textbf{METEOR} & \textbf{CIDEr} \\ 
\hline
Seq2Seq & 18.5 & 38.9 & 18.9 & 1.90 \\
\app-PO & 17.3 & 40.4 & 19.1 & 1.95 \\
\app-TO~($N$=10) & 18.8 & 40.0 & 19.2 & 1.93 \\
\app~($N$=10) & \textbf{19.5} & \textbf{41.6} & \textbf{19.9} & \textbf{2.10} \\
\hline
\end{tabular}}
\end{table}

Further more, we vary the length of code topics $N$ as $3,5,10,15,20$ to analyze the performance change with the increase of the length of code topics.
In fact, we analyze the code topics of training instance in the training set, there are about 18.8 code topics on average which have an assigned probability higher than the minimum threshold (i.e., $10^{-8}$) of topic model.
This is why we set the upper bound to be $20$. 

\begin{figure}[t]
  \centering
  \includegraphics[width=1\columnwidth]{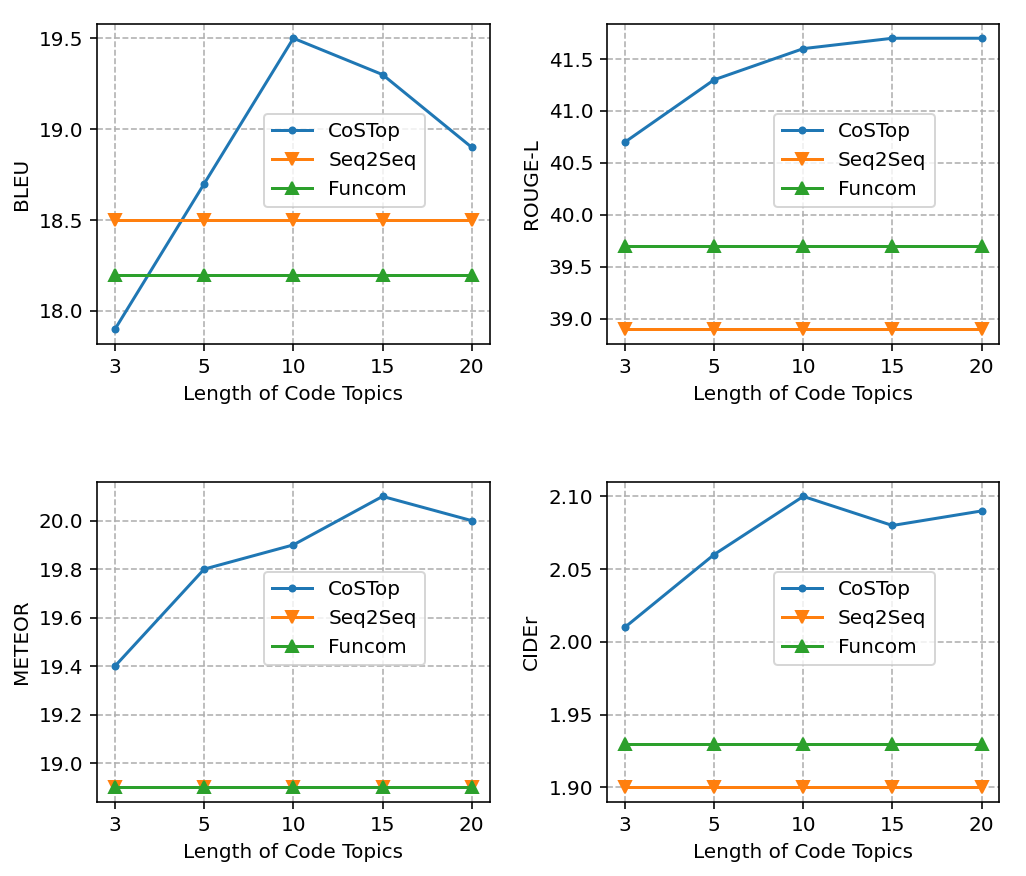}
  \caption{Performance Change with Increase of Length $N$ of Code Topics}
  \label{fig:topics-length-change-performance}
\end{figure}

The change of each performance metric of code topics is shown in Figure~\ref{fig:topics-length-change-performance}.
The observations are as follows.
First, the performance of \app~with more code topics tends to increase when the length is increased from $3$ to $10$ on all the four metrics.
This indicates that the code topics related to code snippets (ranked at the beginning of sorted code topics) could bring positive impact. 
Second, the performance of \app~with more code topics tends to decrease when the length is increased from $15$ to $20$ on most of the metrics (i.e., BLEU, ROUGE-L, and METEOR).
This indicates that the code topics that are only remotely related have negative influence.
Third, the performance reveals a trend of fluctuation with different length of topics in the range from $N=5$ to $N=15$ on different metrics.
Except the BLEU value with code topics length 3, 
the performance of \app~stabilizes at a better position above the baselines on all the metrics. 
This result shows that although the code topics with lower probabilities contain noise that may negatively influence the performance of \app, the code topics related to code snippets contribute a lot to the performance of \app.

\subsection{Human Feedback (RQ3)}\label{sec:rq3}
To evaluate our approach in a setting as close as practice,
we conduct a human study to figure out what human developers think about the automatically generated code summaries.
We randomly select 50 methods whose lines are no less than 5 from the test set and recruit 10 master students for the study.
All the participants major in software engineering and have development experience of more than 4 years.
They are asked to evaluate the code summaries generated by different approaches (i.e., Seq2Seq, Funcom, and \app) together with the golden code summaries (i.e., the original summaries written by human). 
The source code of each method and the class where it resides is provided as the basis.
For each method the four summaries are shown at once in a random order.
Notice that, the participants are not aware of how the code summaries are generated.
For each summary, the participants answer three questions on a 4-points Likert scale (1-Disagree; 2-Somewhat disagree; 3-Somewhat agree; 4-Agree):

\textbf{Expressiveness}. Do you think that the summary is easy to read and understand?

\textbf{Adequacy}. Do you think that the summary contains all the important information?

\textbf{Conciseness}. Do you think that the summary has no (or very little) irrelevant information?

To quantify the evaluation results, different levels of agreement are assigned scores $1, 2, 3, 4$ respectively, and the scores of each code summary evaluated by different participants are averaged to obtain a quantitative score.
The higher the quantitative score is, the better the human feedback is.

\begin{figure}[t]
  \centering
  \includegraphics[width=1\columnwidth]{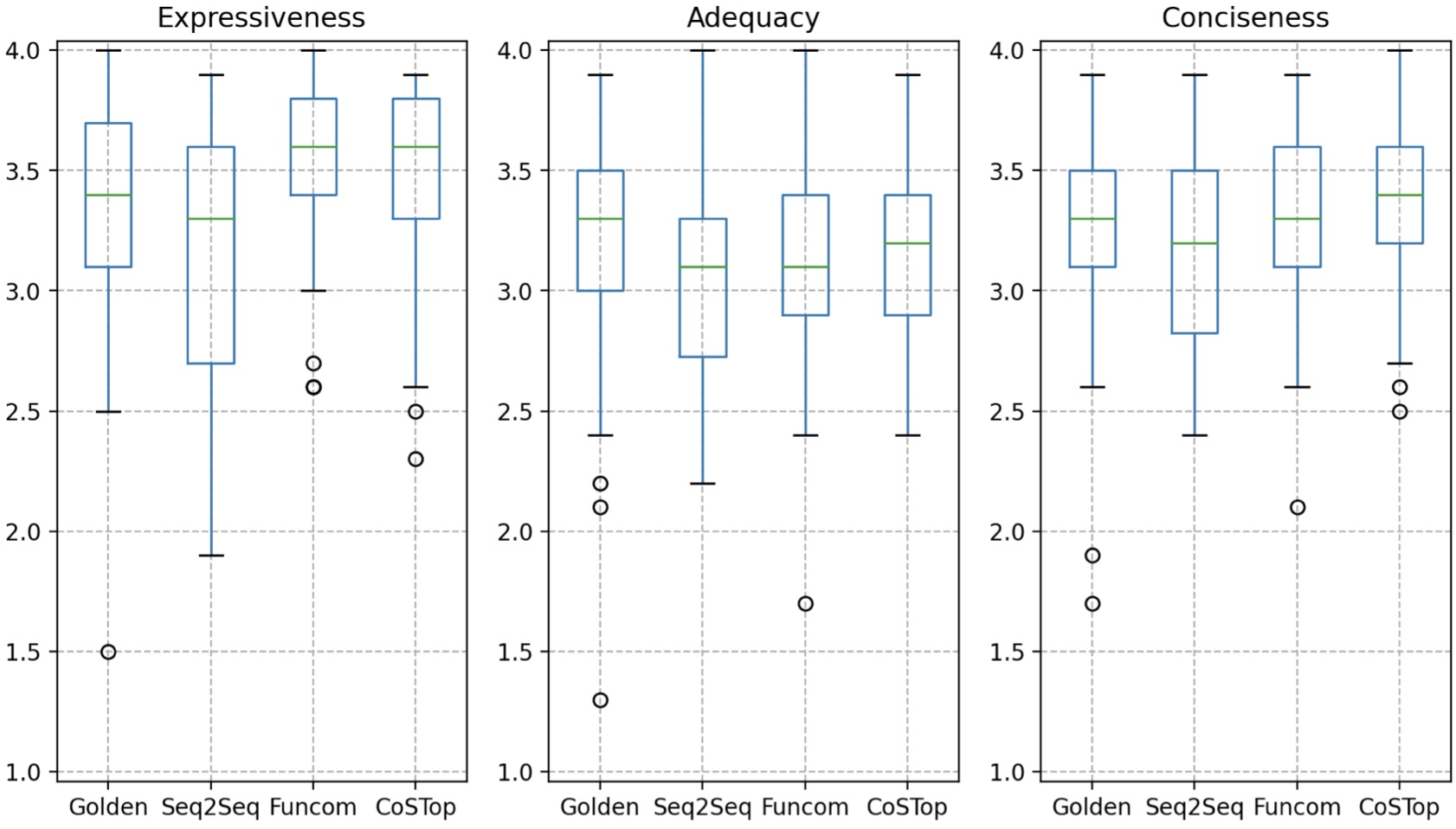}
  \caption{Human Feedback on Generated and Human Written Code Summaries}
  \label{fig:human_study_result_analyse}
\end{figure}

Figure~\ref{fig:human_study_result_analyse} shows the results of human feedback in the form of box plots.
Overall, the feedback is mostly positive (i.e., above 2.5) for all the three criteria of the four kinds of summaries.
The human written summaries tend to contain more information, and therefore have the best feedback on adequacy at the expense of slight decrease in expressiveness and conciseness.
\app~achieves the best feedback on expressiveness and conciseness (even better than golden code summaries), and better feedback on adequacy than Seq2Seq and Funcom.
Comparing with Funcom, \app~achieves comparable feedback on expressiveness (both 3.6 by median) and better feedback on adequacy (3.2 vs. 3.1 by median) and conciseness (3.4 vs. 3.3 by median).

\begin{figure}[h]
  \centering
  \includegraphics[width=0.8\columnwidth]{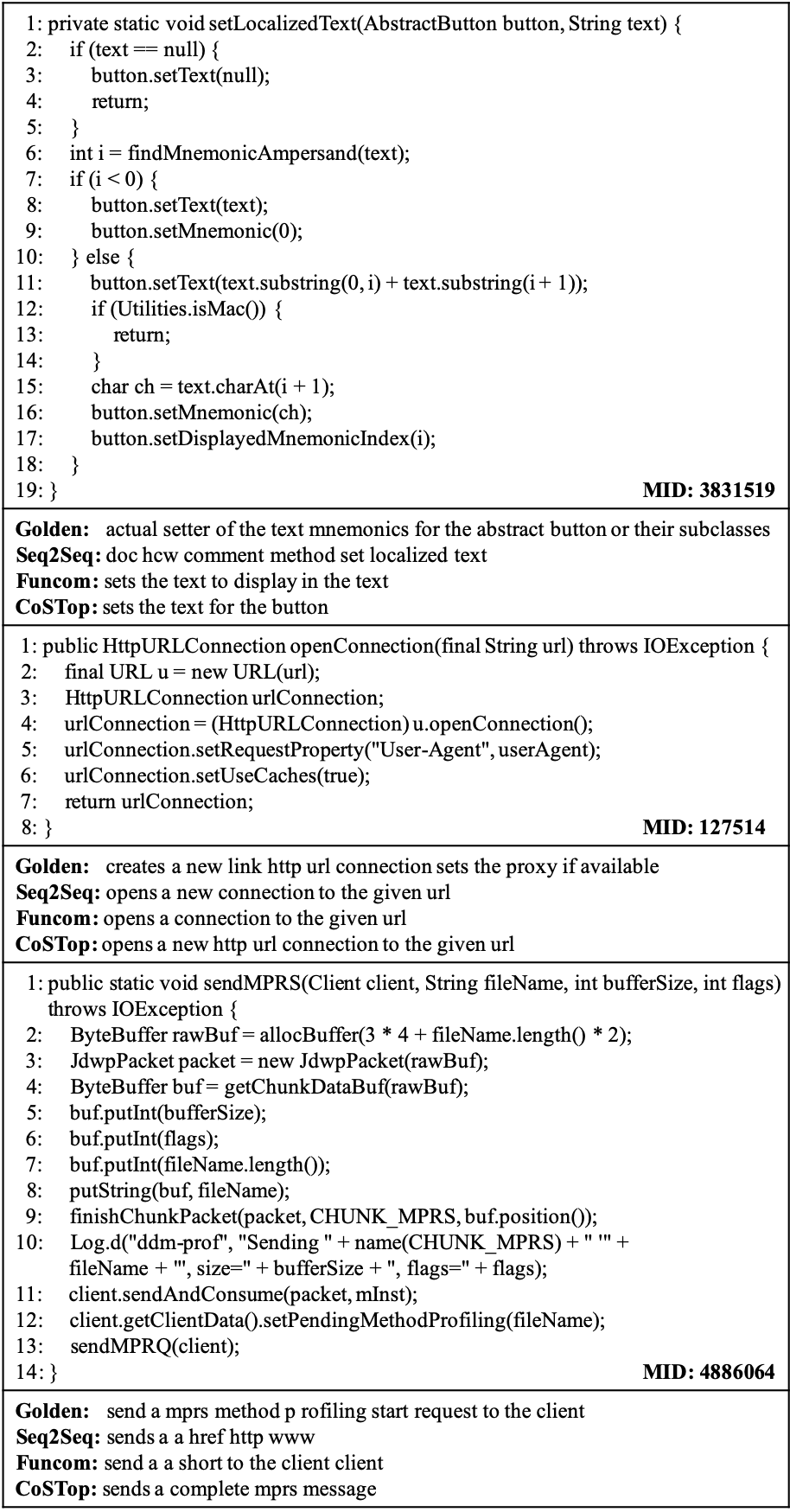}
  \caption{Examples for Qualitative Analysis}
  \label{fig:human_study_example}
\end{figure}

To understand the strength of \app, we manually investigate the methods on which \app~obtains better feedback than the two baseline approaches on at least one of the three criteria. 
Our findings are summarized below and the related examples are shown in Figure~\ref{fig:human_study_example}, each with a method ID (MID) in the dataset.

\textbf{1. Human written summaries and machine generated summaries complement each other.}
Human written summaries tend to convey rationales that may be missed by others, while machine generated summaries tend to summarize the internal processing of code. 
For example, the code snippet with MID 3831519 sets the text for abstract button.
The human written summary highlights the points of actual setter and abstract button with subclasses, and thus is more complex.
In contrast, the summary generated by \app~is more concise. 
Note that the summaries generated by Seq2Seq and Funcom are illegible or incomplete. 
This example suggests that human written summaries and machine generated summaries may complement each other i.e., describing the code from different aspects.
This finding confirms the value of machine generated code summaries.

\textbf{2. Copying helps to generate domain-specific comments.}
Domain specific words exist in nearly every project.
For example, the word ``http'' in the code snippet with MID 127514 is a domain specific word which refers to a protocol in the client/server computing model.
The code snippet opens an http connection to the given URL.
With the ability to copying words, \app~generates the code summary containing the word ``http'' as the protocol of connection.
This makes the code summary closer to the human written one and more adequate.
Another example is the word ``mprs'' in the code snippet with MID 4886064.
It is a domain specific word which means ``Method PRofiling Start''.
The code summaries generated by Seq2Seq and Funcom capture the action of sending something, but cannot generate the word ``mprs'' due to the scarcity of such specific word.
With the copying ability, \app~generates the code summary which includes the word ``mprs'' in the description.

\textbf{3. Relevant topics help bring in related concepts.}
It is challenged for machine generated code summaries to describe the concepts that are implied in the method.
For example, for MID 4886064 the code summaries generated by Seq2Seq and Funcom bring in irrelevant and unnecessary concepts like ``href'', ``http'', ``www'', and ``short''.
Guided by the code topics of the class (e.g., data buffering and message handling), \app~generates the code summary with the concept ``message'' captured.

\subsection{Threats to Validity}
Major threats to the internal validity lie in the following aspects.
First, our approach involves a large code corpus that is automatically generated and the data preprocessing involved in the process may bring errors.
To alleviate the threat, we have followed the same process of data preprocessing in related works.
Second, we do not conduct cross validation in the evaluation and a different choice of split may result in different performance.
The expensive computation cost to make cross validation on such large dataset is a common challenge among deep learning experiments, as it requires a lot of time to training a model (4-8 hours per epoch for different models in this work).
To alleviate the threat, we conduct our evaluation on a large dataset which includes 4,203,565 methods with related class information and split the dataset randomly by project to ensure the objectivity.

The threats to the external validity mainly lie in the fact that our experiments are conducted only on Java code and our own dataset.
Although Java is a popular programming language, the results may not be generalized to other programming languages.
And the same code summarization approach may behave differently on different datasets.
To alleviate the threat, we have tried our best to construct a large enough dataset and made it available online~\cite{ReplicationPackage}.

\section{Related Work}
In recent years, there have been extensive research which are related to automatic code summarization, e.g., summarization generation for methods~\cite{ASE2010TowardsCommentsForJavaMethods,
ICSE2011HighLevelActionMethod,
ICPC2014AutomaticDocumentGenViaContext,
SANER2017ObjectRelatedDescription,
WCRE2010TextSummarizationForSourceCodeIR,
ICPC2014CODES,
SANER2015CloCom,
ACL2016CODENN,
ICML2016ConvolutionalAttentionNetMethodName,
AAAI2018LiangCodeGRU,
ICPC2018Deepcom,
IJCAI2018SummCodeTransferredAPI,
ASE2018DeepReinforcement,
ICLR2019Code2Seq,
ICSE2019LeClair}, summarization generation for classes~\cite{ICPC2013JSummarizerJavaClasses}, and commit message generation~\cite{ASE2017JiangCommitMsgGenNMTDiff,ASE2018HowFarAreWe,MSR2019GenCommitMsgPointerGen}.
Our work focuses on the code summarization for methods based on the source code.
The above-mentioned existing approaches can be categorized into three groups, i.e., template-based approaches, information retrieval (IR) based approaches and deep learning based approaches.

Template-based approaches~\cite{ASE2010TowardsCommentsForJavaMethods, ICSE2011HighLevelActionMethod, ICPC2014AutomaticDocumentGenViaContext, SANER2017ObjectRelatedDescription} focus on mining or designing templates which are then used for code summarization.
These approaches are limited by the quality and variety of the templates that are either mined or manually designed. 

IR based approaches~\cite{WCRE2010TextSummarizationForSourceCodeIR, ICPC2014CODES, SANER2015CloCom} apply IR techniques 
or related techniques (e.g., clone detection) to match code and comments, and then generate comments for some given code.
These approaches compute the similarities between code snippets, and then generate the summary of a given code snippet based on the comments of similar code snippets.
Thus, the effectiveness of such approaches depends on whether similarities between code snippets are computed precisely (which depends on the quality of keywords in source code) and the existence of the similar codes (and its descriptions) in the dataset, both of which are questionable.

Recently, the data-driven approaches, noticeably deep learning based approaches, became popular as they can learn features automatically. Most existing deep learning based approaches regard the code summarization problem as a translation task which translates source code into its summary.
Iyer \emph{et al.}~\cite{ACL2016CODENN} proposed CODE-NN, an end-to-end code summarization generation model for C\# or SQL code snippets based on LSTM networks with attention mechanism.
Allamanis \emph{et al.}~\cite{ICML2016ConvolutionalAttentionNetMethodName} proposed a convolutional attention network to generate the short name-like summaries of code.
Liang \emph{et al.}~\cite{AAAI2018LiangCodeGRU} proposed Code-GRU, a recurrent neural network for generating descriptive comments for code blocks.
Hu \emph{et al.}~\cite{ICPC2018Deepcom} proposed DeepCom, an attentional Seq2Seq model for generating comments for Java methods.
Their model is based on the SBT method, which represents the AST in the form of a sequence which retains the structural information.
Hu \emph{et al.}~\cite{IJCAI2018SummCodeTransferredAPI} proposed a code summarization approach for Java methods based on transfer learning with the assistant of API knowledge.
Wan \emph{et al.}~\cite{ASE2018DeepReinforcement} proposed a deep reinforcement learning based model to generate code summaries for Python methods. 
Alon \emph{et al.}~\cite{ICLR2019Code2Seq} use the compositional paths in AST to represent code snippets and generate comments and method names.
LeClair \emph{et al.}~\cite{ICSE2019LeClair} extended the SBT approach proposed in~\cite{ICPC2018Deepcom} such that both structural and semantic information are taken into account.
Zhang \emph{et al.}~\cite{ICSE2020RetrievalBased} combined neural and retrieval-based techniques for source code summarization.

Our approach differs from the above approaches in two important ways. First, we solve the problem of context-specific OOV by introducing a pointer mechanism to copy tokens directly from source code. Second, we argue the importance of categorizing code according to their topics and introduce a code topic model to guide the generating and copying process. Both have been shown to be effective in solving problems in existing approaches.

In addition, our approach is related to the following work which applied different code representations or neural models for code summarization.
 Zhang \emph{et al.}~\cite{ICSE2019ZhangASTRepresentation} proposed an AST-based neural network representation method for source code, and applied it to two code comprehension tasks, i.e., code classification and clone detection.
LeClair \emph{et al.}~\cite{ICPC2020LeClairGNN} used graph-based neural architecture to improve source code summarization.
Ahmad \emph{et al.}~\cite{ACL2020TransformerBased} applied Transformer model for source code summarization.
Our topic guided pointer-generator model is orthogonal to and can be combined with different code representation methods. We leave it for future work to empirically examine how effective our model is when integrated with different code representation methods.

\section{Conclusion}

In this paper, we propose a neural network model named \model~for code summarization, which uses the topics in a broader context (e.g., class) to guide the neural networks that combine the generation of new words and the copy of existing words in code.
Based on the model we present an approach for generating natural language code summaries at the method level (i.e., method comments).
Our evaluation shows significant improvement over two baseline approaches.
In the future, we will further improve our approach and explore its application in more program comprehension tasks.



%





\ifCLASSOPTIONcaptionsoff
  \newpage
\fi



\bibliographystyle{IEEEtran}
\bibliography{IEEEabrv,reference}


%

%





\end{document}